\documentclass[fleqn,usenatbib]{mnras}

\usepackage{newtxtext,newtxmath}
\usepackage{bm}
\usepackage{makecell, multirow,diagbox}
\usepackage{array}
\usepackage{booktabs}
\usepackage{amsmath} 
\usepackage{orcidlink}
\usepackage{color}
\usepackage{graphicx}                                                           
\usepackage{float}

\usepackage[T1]{fontenc}

\DeclareRobustCommand{\VAN}[3]{#2}
\let\VANthebibliography\thebibliography
\def\thebibliography{\DeclareRobustCommand{\VAN}[3]{##3}\VANthebibliography}

\usepackage{graphicx}	
\usepackage{amsmath}	

\title[SMBH binary candidate PKS J2134-0153: Periodic variability
and time lags]{SMBH binary candidate PKS J2134-0153: Possible multi-band periodic variability and inter-band time lags}

\author[G.-W. Ren et al.]{
Guo-Wei Ren \orcidlink{0000-0002-1497-8371} $^{1}$,
Mouyuan Sun \orcidlink{0000-0002-0771-2153} $^{1}$\thanks{E-mail: msun88@xmu.edu.cn},
Nan Ding \orcidlink{0000-0003-1028-8733} $^{2}$,
Xing Yang $^{3}$, 
and Zhi-Xiang Zhang \orcidlink{0000-0002-2419-6875} $^{1}$
\\
$^{1}$Department of Astronomy, Xiamen University, Xiamen, Fujian 361005, China\\
$^{2}$School of Physical Science and Technology, Kunming University, Kunming 650214, China\\
$^{3}$School of Physical Science and Technology, Guangxi University, Nanning 530004, China
}

\date{Accepted XXX. Received YYY; in original form ZZZ}

\pubyear{2023}

\usepackage[switch]{lineno} 
\begin{document} %

\label{firstpage}
\pagerange{\pageref{firstpage}--\pageref{lastpage}}
\maketitle

\begin{abstract}
Studying the periodic flux-variation behavior of blazars is vital for probing supermassive black hole binaries and the kinematics of relativistic jets. In this work, we report the detection of the multi-band possible periodic variations of the blazar PKS J2134-0153, including the infrared ($1.6(\pm0.4)\times 10^3$ days) and optical ($1.8(\pm1)\times 10^3$ days). The periods in the infrared and optical bands are statistically consistent with the period in the radio band ($P_{\mathrm{Radio}}$$ = 1760\pm33$ days, obtained from our previous work). Moreover, flux variations in different bands are correlated with evident inter-band time delays, and the time lags of infrared and optical emission with respect to radio emission are $(3.3\pm2.3)\times10^{2}$ days and $(3.0\pm2.3)\times10^{2}$ days, respectively. The cross-correlations indicate a common origin of radio, infrared, and optical emission. The relative positions between emission regions of infrared and optical emission to radio emission are estimated according to the time lags, i.e., $0.37\pm0.26$ pc and $0.33\pm0.26$ pc. The relative distances seem to be quantitatively consistent with the theoretical prediction. 
\end{abstract}




\section{Introduction} \label{sec: intro}

Active galactic nuclei (AGNs) emit strong non-stellar emission in the Universe. Blazars are an important subclass of radio-loud AGNs, whose relativistic jets point towards our Earth. Blazars emit a broad range emission from radio to $\gamma$-ray \citep[e.g.,][]{Urry1995, Ulrich1997}. The central engines of blazars are often difficult to be resolved directly because of their small physical scales and the large distance from Earth (i.e., the high redshift); their corresponding angular scales are too small to resolve by current facilities. The multi-band light variations offer an alternative possibility to resolve the central engine in the time domain \citep[e.g.,][]{Blandford1996, Fan1998}. The flux variations provide valuable insights into various aspects of blazar studies, including structures, physical parameters of supermassive black holes (SMBHs, with masses larger than $10^6\ M_{\odot}$), and emission processes \citep[e.g.,][]{Ulrich1997, Gupta2018}. For instance, the minimum variability timescale (minutes to hours) can effectively constrain the emission-region sizes of blazars. Combined with theoretical models and measured time lags between two different light curves via the reverberation mapping (RM) technique \citep[e.g.,][]{Blandford1982}, one can constrain physical parameters of the relativistic jets, e.g., the magnetic field \citep[e.g.,][]{Tavecchio1998}. 

Supermassive black hole binaries (SMBHBs), which includes two orbiting SMBHs, are fascinating sources. The merging of SMBHBs produces gravitational waves \citep[GW; e.g.,][]{Thorne1976, Haehnelt1994, Jaffe2003, Dou2022} detectable by low-frequency GW experiments (e.g., PTA - \citealt{Hobbs2010}; LISA - \citealt{Amaro2023}). Spatially correlated signals consistent with the ultra-low-frequency GW background are detected by the pulsar timing arrays \citep[e.g.,][]{ Agazie2023, EPTA2023, Reardon2023, Xu2023}. Although searching SMBHBs is attractive, the direct identification of close SMBHBs ($<$ 1 pc) is highly challenging because of the angular resolution limit of current telescopes \citep[e.g.,][]{Dou2022}. 

Searching for periodicity in multi-band light curves is an attractive method for identifying SMBHB candidates. Year-like periodic flux variations can probe the presence of SMBHB systems \citep[e.g.,][]{Begelman1980, Lehto1996, King2013, Ackermann2015, Liu2015, Bon2016, Charisi2016, Severgnini2018, Liu2019, Kovacevic2020, Zhang2022a, Kun2023}. A notable example is OJ 287, which shows a quasi 12-year periodicity in its light curves \citep[e.g.,][]{Valtonen2006, Valtonen2008, Gupta2022, Sinitsyna2022, Valtonen2024}. This periodicity is explained by a SMBHB system where the secondary SMBH crosses the accretion disk of the primary SMBH, leading to the observed quasi-periodic variations \citep[e.g.,][]{Dey2018}. 

The quasi-periodic behavior over year-like timescales can be explained by models with jet precession \citep[e.g.,][]{Romero2000, Stirling2003, Caproni2013} or helical jets \citep[e.g.,][]{Conway1993, Tateyama1998, Mohan2015}. In the SMBHB scenario, the jet precession is driven by the gravitational torque from the secondary black hole in the misaligned orbits and the intrinsic helical jet structure could be driven by a coiled magnetic field \citep[e.g.,][]{Vlahakis2004, Miller2006}. As a result, the viewing angle between the observer and the jet changes periodically, which is responsible for the observed periodic variations \citep[e.g.,][]{Villata1999}. It should be noted that the jet precession may happen in single SMBH system. According to the disk-driven precession model \citep[e.g.,][]{Sarazin1980, Lu1990, Lu2005}, the misalignment between the SMBH’s spin axis and the accretion disk and the Lense-Thirring effect makes the inner disk and the jet precess. The periods produced by this model are often much longer than ten years (more details, see Section \ref{sec:discussion_period}). Hence, it is important to find AGNs that show short-period (less than several years) periodic flux variations. 

A number of works have been devoted to identifying SMBHB candidates via periodic variations. \cite{Graham2015} reported a possible close supermassive black-hole binary in PG 1302-102 with optical periodicity \citep[the period is $P = 1,884 \pm 88$ days, but also see][questioning this source]{Liu2018}. \cite{Sesana2018} selected 111 SMBH binary candidates from approximately 250,000 quasar light curves observed by the Catalina Real-time Transient Survey (CRTS) due to their potentially periodic nature. \cite{Bhatta2018} proposed that the blazar, J1043+2408, is an SMBH binary candidate because of the periodic flux variation. \cite{Li2018} reported a $\sim$ 6.1 year quasi-periodicity of the blazar S5 0716+714, which is also a SMBH binary candidate. \cite{Serafinelli2020} found that the Seyfert 1.5 galaxy Mrk 915 emerges as a candidate SMBHB from the Swift Burst Alert Telescope (BAT) 105-month light curves of 553 AGNs. In our previous work, we reported a $\sim$4.69 years period in the radio-band light curve of BL Lac PKS J2134-0153 (\citealt{Ren2021}; hereafter R21; see also \citealt{ONeill2022}; hereafter ON22), and we speculate that this is a close SMBH binary candidate with a small orbital separation of $\sim 0.006$ pc. 

In this work, we report the multi-wavelength (including the infrared, optical, and $\gamma$-ray) study of PKS J2134-0153. We also find inter-band correlations with evident time delays. Hence, we also constrain the emission-region sizes of PKS J2134-0153. This manuscript is formatted as follows. In Section \ref{sec:data}, we describe the multi-band light curves of PKS J2134-0153. In Section \ref{sec:analysis}, we introduce the data analysis methods. In Section \ref{sec:results}, we present the results. In Section \ref{sec:discussion}, we give the discussions. Conclusions are made in Section \ref{sec:conclusion}.

Throughout this paper, a $\Lambda$CDM cosmological model with $\Omega_{\mathrm{m}} = 0.27$, $\Omega_{\mathrm{\Lambda}} = 0.73$, $H_{\mathrm{0}} = 70\;\mathrm{km\;s^{-1}\;Mpc^{-1}}$ is adopted for calculating the luminosity distance.

\section{Light curves} \label{sec:data}
BL Lac PKS J2134-0153 \cite[$z = 1.285$; e.g.,][]{Rector2001, Truebenbach2017} is observed in various bands, including radio, infrared, optical, and $\gamma$-ray from 2008-2023. The multi-band data we used to investigate the nature of the source are detailed in the following subsections. For gamma-ray data, please refer to Appendix~\ref{sec:gamma}.

\subsection{Radio observations}
\label{sec:radio}
PKS J2134-0153 is observed in the radio band. The radio observations are discussed in detail in our previous work (R21; see also ON22) and will not be re-analyzed in this work. In R21, we perform the Lomb-Scargle Periodogram \citep[LSP;][]{Lomb1976, Scargle1982} and the weighted wavelet Z-transform \citep[WWZ;][]{Foster1996} techniques to analyze the 15 GHz radio light curve obtained by the Owens Valley Radio Observatory (OVRO) 40-m telescope. The cadence and duration of the analyzed light curve are $\sim$ 7 days and $\sim$ 11.4 yrs in the observed frame, respectively. We find that the radio flux variations are periodic with an observed-frame period of $4.69 \pm 0.14$ yrs, and can be fitted well by the sinusoidal function (R21). ON22 also use LSP and WWZ to analyze the observed-frame 45.1 years radio light curve of the same target. Their data are collected by the Haystack Observatory (15.5 GHz, with a cadence of $\sim 1$ month in the observed frame), the OVRO (15 GHz), and the University of Michigan Radio Astronomy Observatory (14.5 GHz, with a cadence of $\sim 3$ months in the observed frame). The period ($2.082\pm0.003$ years in the rest frame or $4.757\pm0.007$ years in the observed frame) obtained by ON22 is close to our result. The radio band light curve and the best-fitting sinusoidal function obtained by R21 are shown in the top panel of Figure \ref{fig:LC}. This periodic feature suggests that PKS J2134-0153 may be a candidate SMBHB. The disk-driven precession model that involves a single SMBH can hardly account for the observed period (see Section \ref{sec:discussion_period}). 

\subsection{Infrared observations}
\label{sec:wise-obs}
PKS J2134-0153 is observed by the Wide-field Infrared Survey Explorer \citep[WISE;][]{Wright2010} and the Near-Earth Object Wide-field Infrared Survey Explorer \citep [NEOWISE;][]{Mainzer2014} in the infrared bands. WISE was launched on 2009-12-14, which is a 40 cm diameter infrared telescope designed to perform a repetitive all-sky survey at four infrared wavelength bands including $W1$, $W2$, $W3$, and $W4$ (centered at 3.4, 4.6, 12, and 22 $\mathrm{\mu m}$ in the observed frame). The survey was shut down on 2010-09-29 because the cooling material was depleted. The WISE spacecraft was reoperated in the all-sky survey model in $W1$ and $W2$ bands on 2013-12-13, which is known as the NEOWISE survey. Each sky region is observed by the WISE telescope every 6 months.  

The infrared band observations of PKS J2134-0153 are obtained from the NASA/IPAC Infrared Science Archive\footnote{\url{https://irsa.ipac.caltech.edu/frontpage/ }} with a maximum matching radius of ${3}''$. For the $W1$ and $W2$ bands, 21 and 239 pointings are obtained by WISE (between 2010-05-15 and 2010-11-11) and NEOWISE (from 2014-05-16 to 2022-11-1), respectively. The time span of these observations is 4553 days ($\sim$ 12.5 yrs) in the observed frame. The observations include 20 epochs, and each spans 1\textendash 7 days. The time interval between two adjacent epochs is roughly 180 days, except for the gap from 2010-09-29 to 2013-12-13. Following \cite{Sheng2017, Sheng2020} and \cite{Li2023}, the bad observations, which were affected by low image quality ('qi$\_$fact' $< 1$), scattered moonlight ('moon$\_$masked' $> 0$), and the satellite’s small separation from the South-Atlantic Anomaly ('saa$\_$sep' $< 5$), are removed. Finally, we retained $162$ data points (the blue and green dots in the second panel of Figure \ref{fig:LC}) in the $19$ epochs. For each epoch, we re-bin the WISE observations because we are interested in the long-term variations. The observations were not considered if the total number of the data points in an epoch is less than three. The binned $W1$ and $W2$ light curves are shown as the black and purple stars in the second panel of Figure \ref{fig:LC}. It can be seen that there are significant flux variations in the two bands, and the variations appear quasi-periodic. 

In ON22, the infrared ($W$1 and $W$2 bands) light curves of this source are obtained from WISE and NEOWISE (up to MJD 59150, i.e., 730 days shorter than our compiled WISE light curves). They did not perform periodic analyses of the infrared light curves. They calculated the cross-correlation of the infrared and OVRO 15 Ghz radio light curves and obtained a $\sim$ 250\textendash350 days time lag between the two light curves (see figure 6 of ON22). We independently measure the infrared-radio time lag because our requested WISE light curves are longer than ON22. Our infrared-radio time lag is statistically consistent with that of ON22 (see Section \ref{sec:timelags}).  

\subsection{Optical observations}
\label{sec:optical}
\begin{figure*}
\includegraphics[width=1 \textwidth]{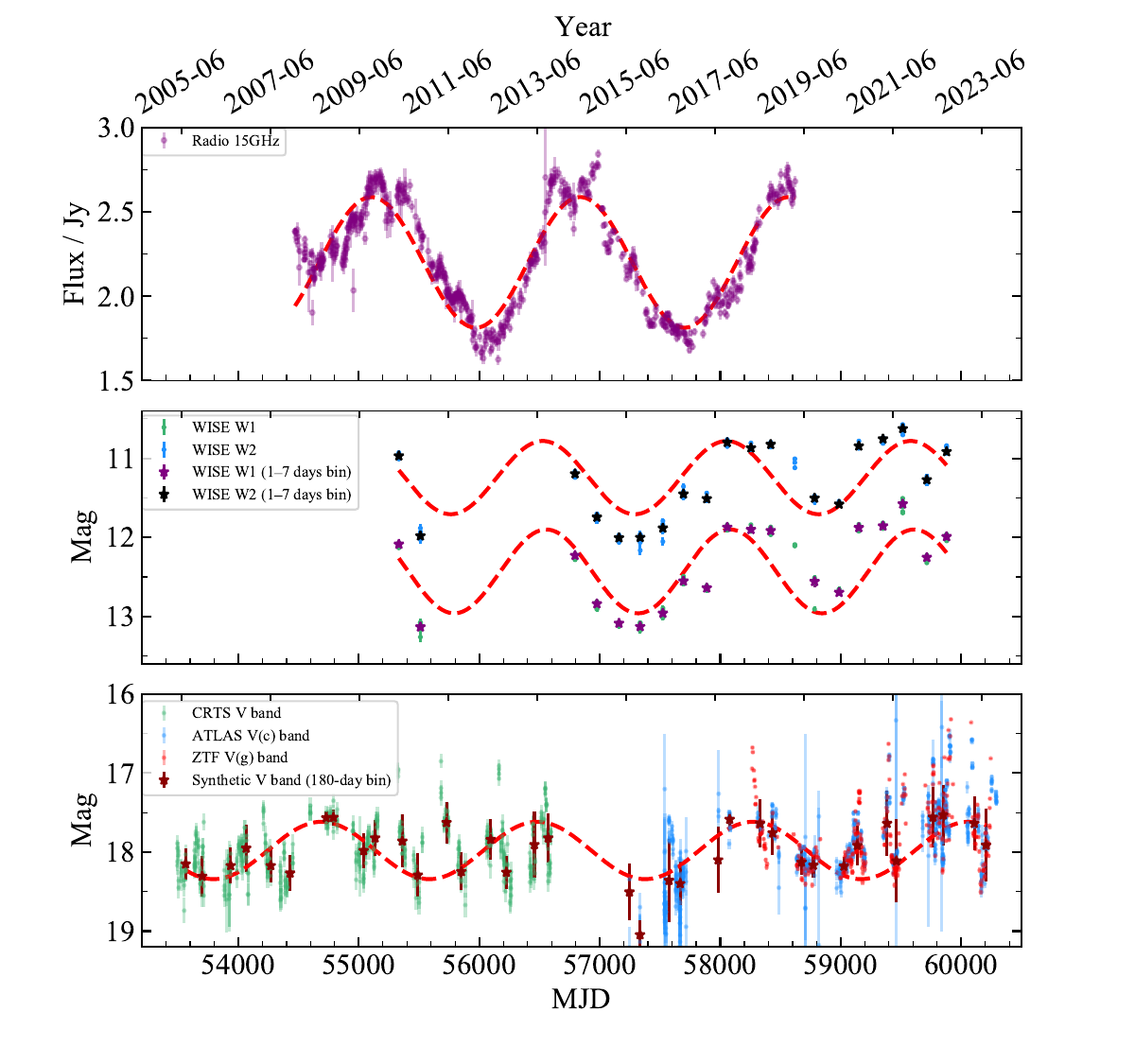}
\caption{The multi-band light curves of PKS J2134-0153. Top panel: the radio light curve and the best-fitting sinusoidal function from R21. The radio light curve shows possible double peaks at the flux maxima, whose origin remains unclear. Middle panel: the WISE infrared light curves in the $W1$ (green circles) and $W2$ bands (blue circles); the purple and black stars represent the $1$--$7$ days re-binned $W1$ and $W2$ light curves. Note that the re-binned points in 2019-05 are not considered since the total number of the data points is less than 3 in this epoch. Bottom panel: the synthetic $V$ band light curve is composed of CRTS $V$ band (green circles), ATLAS $c$ band (blue circles), and ZTF $g$ band (red circles). The dark red stars represent the 180-day binned $V$ band light curve. The dashed red curves show the best-fitting sinusoidal functions of the binned WISE infrared and the binned $V$ band light curves. There are evident flux variations in each band.
\label{fig:LC}}
\end{figure*}

The light curves in the optical bands include the $V$, $g$, $r$, $i$, and $c$ bands. The $V$-band observations are obtained by the Catalina Real-time Transient Survey\footnote{\url{http://crts.caltech.edu/}} \citep[CRTS,][]{Drake2009}, from 2005-05-03 to 2013-10-27 (MJD 53494 to MJD 56593), spanning 3099 days ($\sim$ 8.5 years) in the observed frame, and have a total of 415 data points (the green points in the bottom panel of Figure \ref{fig:LC}). The $g$, $r$, and $i$ band observations are obtained from the Zwicky Transient Facility\footnote{\url{https://www.ztf.caltech.edu/}} \citep[ZTF,][]{Bellm2019, Masci2019}. The start and end time of the $g$, $r$, and $i$ band light curves (see Figure \ref{fig:LC_optical}) are from MJD 58263 to MJD 60244 (1981 days), from MJD 58256 to MJD 60244 (1988 days), and from MJD 58280 to MJD 58771 (491 days), respectively. The corresponding total number of data points is 326, 416, and 42, respectively. The observations in the $c$ band are carried out with the Asteroid Terrestrial-impact Last Alert System\footnote{\url{https://atlas.fallingstar.com/}} \citep[ATLAS,][]{Heinze2018, Tonry2018}, from MJD 57246 to MJD 60292, spanning 3046 days ($\sim 8.3$ years) in the observed frame, and has a total of $611$ data points. The orginal optical light curves are shown in Figure \ref{fig:LC_optical}.

We adopt the CRTS, ZTF, and ATLAS data to construct a $6798$-day (observed frame) long synthetic $V$-band light curve as follows,
\begin{itemize}
\item The optical spectrum (plate-mjd-fiberid=4384-56105-0988, shown in Figure \ref{fig:spectra}) of PKS J2134-0153 was obtained from Sloan Digital Sky Survey (SDSS) DR16 \citep[][]{Ahumada2020}. The filter response curves of CRTS-$V$ band, ZTF-$g$ band, and ATLAS-$c$ band are obtained from the Filter Profile Service.\footnote{\url{http://svo2.cab.inta-csic.es/theory/fps/}} 

\item We convolve the SDSS spectrum with the aforementioned filter response curves to obtain the corresponding magnitudes for CRTS $V$ band ($m_{\mathrm{SDSS},\ \mathrm{CRTS}-V}$), ZTF $g$ band ($m_{\mathrm{SDSS,\ ZTF}-g}$), and ATLAS $c$ band ($m_{\mathrm{SDSS,\ ATLAS}-c}$). 

\item We calculate the magnitude differences, $\Delta m_{\mathrm{SDSS,}\ g-V} =m_{\mathrm{SDSS,\ CRTS-}V} - m_{\mathrm{SDSS,\ ZTF-}g}=-0.21$ mag and $\Delta m_{\mathrm{SDSS,\ }c-V} =m_{\mathrm{SDSS,\ CRTS-}V}\ -\  m_{\mathrm{SDSS,\ ATLAS-}c}=-0.04$ mag. 

\item We add $\Delta m_{\mathrm{SDSS,\ gV}}$ and $\Delta m_{\mathrm{SDSS,\ cV}}$ to the original ZTF $g$ and ATLAS $c$ light curves, respectively.

\item The offset-corrected ZTF $g$ and ATLAS $c$ light curves and the CRTS-$V$ band light curve are appended to build a new synthetic $V$-band light curve. 
\end{itemize}
The new synthetic $V$-band light curve has 1352 data points from 2005-05-03 to 2023-12-14 (MJD 53494 to MJD 60292, see the bottom panel of Figure \ref{fig:LC}). Similar to the infrared band light curves, we re-bin the new synthetic $V$-band light curve with 180-day bin (the dark red stars in the bottom panel of Figure \ref{fig:LC}). It is worth noting that ON22 have previously analyzed the CTRS data (approximately from MJD 53491 to MJD 57357) and part of the ZTF data (approximately from MJD 58239 to MJD 59458) and did not find any periodicity. Here, we build longer light curves with better cadences than ON22 from the CTRS, ZTF, and ATLAS datasets. For a detailed periodicity analysis, see Section \ref{sec:period}.

\section{Data analysis} \label{sec:analysis}

We search the periodic variation in the multi-band light curves of PKS J2134-0153 via the Lomb-Scargle Periodogram (LSP) \citep{Lomb1976, Scargle1982}. We use the damped random walk (DRW) simulations to estimate the significance of the period. In addition, we use the Cross-Correlation Function (CCF) to calculate the inter-band time lags.

\subsection{Lomb-Scargle Periodogram} \label{sec:LSP}
The LSP is a commonly used method in time series analysis, which is based on the Discrete Fourier transform (DFT) method \citep[e.g.,][]{Bhatta2018, Prince2023}. The LSP is suitable for detecting and characterizing periodic variations in unevenly sampled time series, and avoiding the interpolation of data gaps \citep[e.g.,][]{Li2017, Bhatta2018, VanderPlas2018}. In this analysis, we run the period search on a frequency grid. To construct the frequency grid, one must specify the frequency limits ($f_{\mathrm{min}}$, $f_{\mathrm{max}}$) and the grid spacing (determined by the total number of required periodogram evaluations, $N_{\mathrm{eval}}$). According to \cite{VanderPlas2018}, the minimum frequency is adopted as $f_{\mathrm{min}} = 1/t_{\mathrm{span}}$, where $t_{\mathrm{span}}$ is the time span of the light curve. The maximum frequency is adopted as $f_{\mathrm{max}} = 1/(2\left \langle \delta t \right \rangle)$, where $\left \langle \delta t \right \rangle$ represents the mean sampling time ($5$ days) of the synthetic V-band light curve. The total number of required periodogram evaluations is
\begin{equation}
N_{\mathrm{eval}}=n_{\mathrm{0}}t_{\mathrm{span}}f_{\mathrm{max}} \label{equ:N_eval},
\end{equation}
and we adopt $n_{\mathrm{0}}$=10 (e.g., \citealt{Urry1995}; \citealt{Debosscher2007}; \citealt{Richards2012}; R21). Frequencies are linearly and uniformly distributed. The light curves are fitted by the following sinusoidal function 
\begin{equation}
y\left ( t, A, \varphi, C \right ) = A\sin \left ( 2\pi f_{\mathrm{best}}t+\varphi   \right )+C  \label{equ:model}  ,
\end{equation}
where $t$ is the sequence of observation times, $f_{\mathrm{best}}$ is the best frequency (corresponding to the maximum power) fixed by the LSP method, and $A$, $\varphi$, and $C$ represent the amplitude, phase, and offset of the simulated light curve, respectively. The best-fit model parameters are obtained via the standard minimization $\chi^{2}$ method \citep[LMFIT;][]{Newville2004}. Then, for each band, the best-fitting sinusoidal function can be obtained from the best-fitting parameters (i.e., the dashed red curves in Figure ~\ref{fig:LC}).

\subsection{Significance estimation} \label{sec:Significance}
The significance levels of an LSP power should be carefully estimated. The red noise can easily mimic periodic signals for unevenly sampled light curves and cause large false peaks in the LSP power. Meanwhile, AGN light curves can be well described by the DRW \citep[a popular red noise model;][]{Kelly2009, Kozlowski2010, MacLeod2012, Zu2013, Secunda2024}, whose power spectral density (PSD) is 
\begin{equation}
P(f) = C / (1 + (f/f_{\mathrm{break}})^2) \\,
\end{equation}
The free parameters are the normalization ($C$) and the breaking frequency ($f_{\mathrm{break}}$). The PSD of the DRW process is dominated by white noise at the low-frequency end and a random walk at the high-frequency end. The DRW model has been applied to fit the light curves of blazars \citep[e.g.,][]{Zhang2022b}. We use {\tt celerite} \citep[][]{Foreman2017} code to fit the DRW model to each light curve and obtain the best-fitting parameters (i.e., $f_{\mathrm{break}}$ and $C$) by maximizing the corresponding {\tt celerite} likelihood \citep[via {\tt scipy};][]{Virtanen2020}. Second, we use {\tt celerite} to generate 20, 000 evenly sampled light curves according to the best-fitting DRW parameters, and the time duration of each one is the same as the observed light curve. For each of the 20, 000 generated light curves, we then use the linear interpolation method to construct a light curve that shares the same sampling as the observed one. That is, like real observations, each simulated light curve is unevenly sampled. Third, the simulated 20, 000 unevenly sampled light curves are analyzed by the LSP method, and 20, 000 simulated LSP power spectra are obtained. Again, just like the PSDs of real light curves, the PSD of each simulated light curve is affected by the same window function and uneven sampling. Finally, we calculate the 84.13-th, 97.72-th, and 99.87-th percentiles of the 20, 000 simulated LSP power for each frequency, which correspond to the $1\sigma$, $2\sigma$, and $3\sigma$ significance levels. The periodic analysis of the radio observations, which suggests that the period is $1760\pm 33$ days (R21; ON22), provides important prior knowledge of the expected period in the optical and IR bands. Hence, we only focus on the period range of $1661$--$1859$ days (i.e., the $3\sigma$ lower and upper limits of the radio period) and do not need to consider the "Look Elsewhere" effect \citep[e.g.,][]{Gross2010}. The corresponding period is insignificant if the observed LSP power peak is lower than the simulated LSP power spectra. Our code to perform the significance analysis can be freely downloaded from the supplementary material (Section~\ref{sec:download}). 

\subsection{Cross Correlation Function \label{sec:CCF}}
We calculate the possible time lags between the multi-band light curves of PKS J2134-0153 via the public Python version of the interpolated cross-correlation function \citep[ICCF;][]{Peterson1998, Peterson2004}, \texttt{PyCCF} \citep[][]{Sun2018}, which is one of the commonly used methods in time series analysis of AGNs. The code uses the linear interpolation method to deal with the unevenly sampled light curves, and calculate the cross-correlation coefficient as a function of the time lag for two light curves. The ICCF is evaluated for the time lag ($\tau$) range from $-4000$ days to $4000$ days. The searching step $\Delta \tau$ should be smaller than the median sampling time $\left \langle \delta t \right \rangle$ of the light curves. We adopt $\Delta \tau =60$ days in this work. As will be demonstrated in Section \ref{sec:timelags} and shown in Figure ~\ref{fig:CCF}, the ICCF has two strong peaks. For each peak, we calculate its corresponding centroid only using the ICCF for the time lags around the peak (see Table~\ref{tab:paramenters_CCF}). In this work, we adopt the centroid of the cross-correlation function ($\tau_{\mathrm{cent}}$) using only time lags with $r > 0.8 r_{\mathrm{max}}$, where $r_{\mathrm{max}}$ represents the peak value of the CCF. The 1$\sigma$ confidence ranges can be calculated via the popular random subset selection and flux redistribution method. \texttt{PyCCF} adopts the student's $t$ statistic to calculate the $p$-value of each peak; the $p$-value is the probability of observing similar or stronger cross-correlations under the no-correlation hypothesis. If the $p$-value is larger than 0.01, the peak is statistically insignificant. 

\section{Results} \label{sec:results}
\subsection{Lomb-Scargle results} \label{sec:period}

We perform the LSP analysis of the light curves in the infrared and $V$ bands of PKS J2134-0153. We compute their LSPs from the minimum ($f_{\mathrm{min}}$) to the maximum frequencies ($f_{\mathrm{max}}$). The values of $f_{\mathrm{min}}$ and $f_{\mathrm{max}}$ for each light curve are obtained according to the description in Section \ref{sec:LSP}. The total number of periodogram frequencies $N_{\mathrm{eval}}$ are set according to Equation \ref{equ:N_eval}. $f_{\mathrm{min}}$, $f_{\mathrm{max}}$, and $N_{\mathrm{eval}}$ for each light curve are shown in Table \ref{tab:paramenters}.  

\begin{figure*}
\includegraphics[width=0.4 \textwidth]{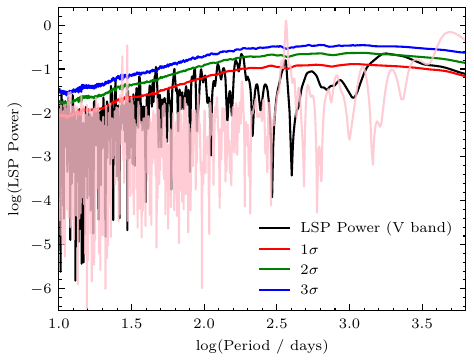}
\includegraphics[width=0.42 \textwidth]{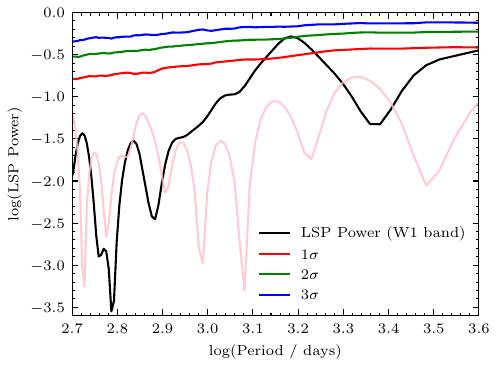}
\caption{The Lomb-Scargle periodograms of the $V$ band (left panel) and WISE $W1$ observations (right panel) of PKS J2134-0153. The red, green, and blue solid curves represent the $1\sigma$, $2\sigma$, and $3\sigma$ significance levels of the DRW mock light curves (i.e., do not contain periodic signals). The pink curve corresponds to the Lomb-Scargle periodogram of the irregular sampling. The significance level of the highest peak at $1.8\times 10^3$ days for the V-band light curve is $2.5\sigma$. } \label{fig:LSP}
\end{figure*}

The LSP power spectra in the synthetic unbinned $V$ band are shown in the left panel of Figure \ref{fig:LSP}. Some ``fake'' PSD peaks are likely caused by the irregular sampling (the pink curves in Figure \ref{fig:LSP}). We test the significance of peaks using the method described in Section \ref{sec:Significance}. There are several peaks that cannot be entirely attributed to the irregular sampling effects. All of these peaks have significance levels of $\lesssim 2.5\sigma$. The significance level is based on the fact that the possibility of the DRW process reproducing the observed peak, or a more statistically significant peak, is $1\%$. The peaks with periods of $\sim 100$ days are not observed in the radio data and are neglected in subsequent analysis. The highest peak corresponds to the period $P_{\mathrm{V}} = 1.8(\pm 1)\times 10^3$ days and seems not to be able to be explained by the irregular sampling. This period is very close to the period detected in the radio light curve (Section~\ref{sec:radio}). Hence, we will focus on this tentative detection in the $V$-band. It is worth noting that ON22 have performed a similar periodic analysis. Unlike our work, ON22 only consider the CRTS and ZTF observations, with 322 data points and the baseline of $\sim$ 6000 days. Hence, they did not find any significant peak in their PSD. We think that the ATLAS data can fill the gap between CRTS and ZTF observations, hence improving the PSD estimation. Indeed, if we neglect the ATLAS observations and the new ZTF data released in 2022 and 2023 in our synthetic $V$ band light curve, we also cannot detect any statistically significant periods. Our synthetic 180-day bin $V$ band light curve is fitted with a sinusoidal function (i.e., Equation \ref{equ:model}, and $f_{\mathrm{best}}$ is fixed to $1/P_{\mathrm{V}}$) and the best fitting parameters are $A_{\mathrm{V}} = 0.36$, $\varphi_{\mathrm{V}} = 1.3$, and $C_{\mathrm{V}} = 17.97$, respectively. The best-fitting curve is shown in the bottom panel of Figure \ref{fig:LC}. The $\chi^{2}$ of the sinusoidal fit is 51. We also fit the data with a constant function, and the $\chi^{2}$ is 123. In addition, we calculated the Akaike Information Criterion (AIC) for the sinusoidal and constant fits to be 19 and 46, respectively. Statistically speaking, if the AIC difference between two models is larger than $10$, the model with a larger AIC value has relatively little support \citep[e.g.,][]{Burnham2011}. Hence, the sinusoidal function is preferred over the constant fit. It should be noted that the AIC difference does not appear to be large enough for the sinusoidal fit to completely dominate.

The IR variations appear to be periodic. The LSP power spectrum for the unbinned $W1$ light curve is shown in the right panel of Figure \ref{fig:LSP}. The LSP power spectrum suggests that there is also a $P_{\mathrm{IR}} = 1.6(\pm0.4)\times 10^3$ days period in the IR light curve. However, the confidence of this peak is about 2$\sigma$ due to the small number of IR data points. We performed the same sinusoidal fit ($f_{\mathrm{best}}$ is fixed to $1/P_{\mathrm{IR}}$) to the binned infrared ($W1$) light curve (the middle panel of Figure \ref{fig:LC}). The best fitting parameters are $A_{\mathrm{IR}} = -0.53$, $\varphi_{\mathrm{IR}} = 1.27$, and $C_{\mathrm{IR}} = 12.43$. The $\chi^{2}$ of the sinusoidal fit is 2400, and the $\chi^{2}$ of the constant function fit is 5067 (the same fitting analysis was performed for the $W2$ band; see the middle panel of Figure \ref{fig:LC}). Similarly, we calculate the AIC values of 97 and 108 for the sinusoidal and constant fits of the W1 band light curve, respectively. Thus, the light curve in the IR band can be better fitted by the sinusoidal function \citep[e.g.,][]{Burnham2011}{}{}. The $1.6(\pm0.4)\times 10^3$ days is statistically consistent with the period in the radio band ($P_{\mathrm{Radio}}$$ = 1760\pm33$ days; R21; ON22) and optical band. Hence, it is very likely that the same fluctuating synchrotron radiation produces radio, infrared, and optical variations. 

\begin{figure*}
\includegraphics[width=0.4 \textwidth]{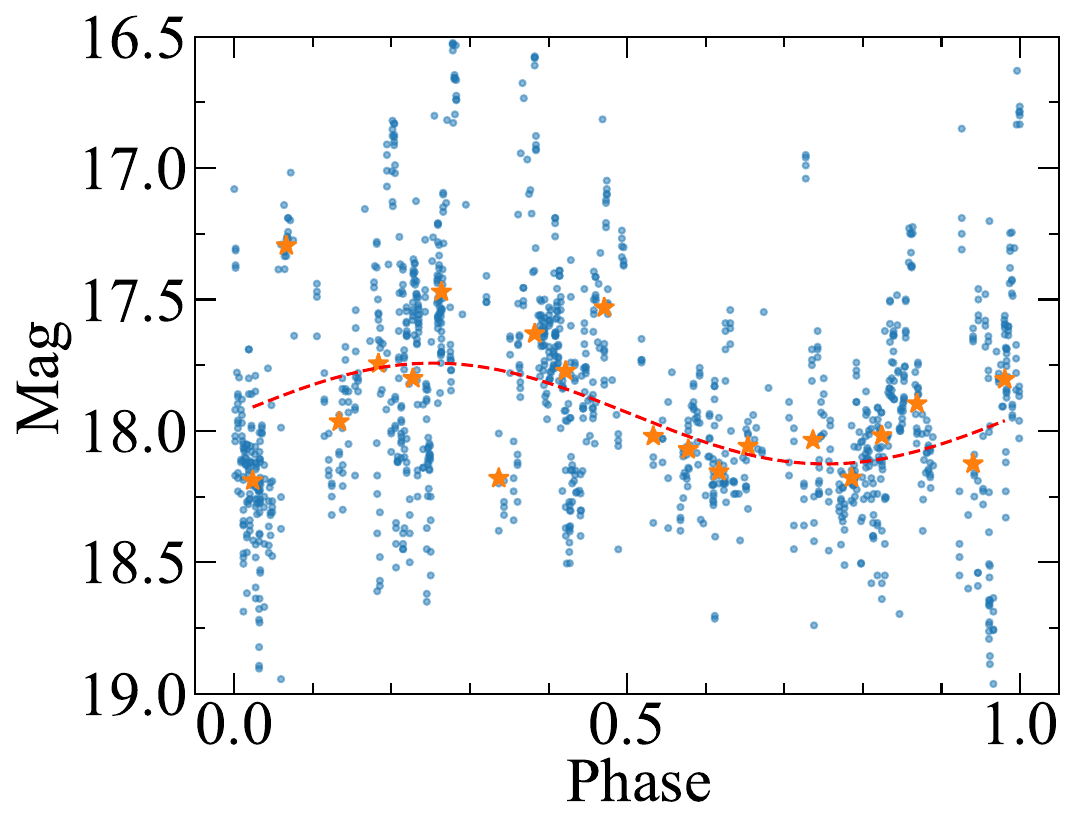}
\includegraphics[width=0.42 \textwidth]{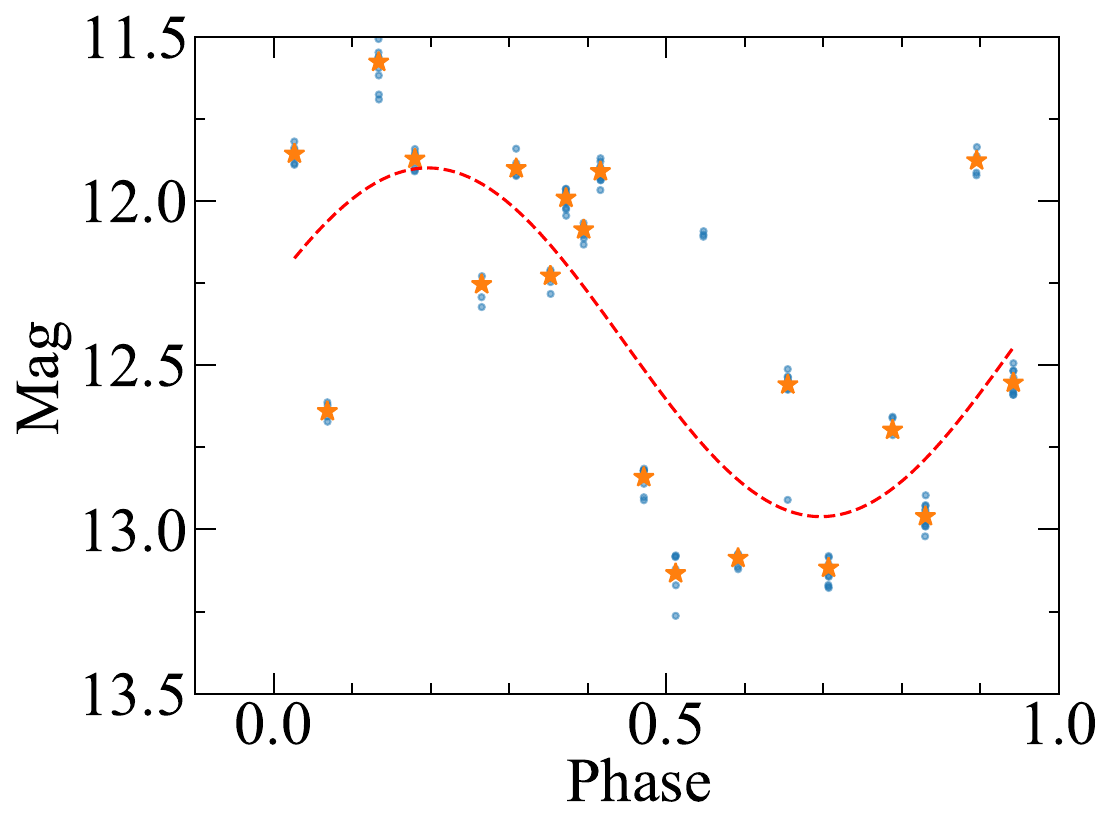}
\caption{Phase-folded $V$-band and IR $W1$ light curves (blue circles). The dashed curves represent the best-fitting sinusoidal functions. The orange stars represent phase-folded binned light curves. \label{fig:folding}}
\end{figure*}

We phase-fold the $V$ band and IR light curves according to their LSP spectral peaks. First, the phase is calculated as $((\mathrm{MJD-MJD_{ref}})\%P)/P$, where $\%$ and $P$ represent the modulo operator and period, respectively. Second, the light curves are rearranged according to the phases. Third, rearranged light curves are re-bin with a phase bin size of $0.05$ for the optical data. For the WISE IR bands, we re-bin the phase light curves by the WISE observed epoch. The results are shown in Figure \ref{fig:folding}. It appears that the light curves show sinusoidal variations on top of some short-term random variations. In addition, the phase-fold $V$ and $IR$ band light curves are fitted by the sinusoidal function and constant function. We calculate the delta AIC values 187 (for the $V$-band) and 60 (for the IR-band). Hence, the constant model is less supported by the data than the sinusoidal model \citep[e.g.,][]{Burnham2011}. Compared with the radio variations, the sinusoidal variations are less prominent in the phase-folded optical light curve, possibly because the accretion disk also produces a significant fraction of optical emission and this disk component varies statistically as commonly seen in radio-quiet AGNs \citep{Ulrich1997}. 

Overall, possible periodic signals are present in PKS J2134-0153 optical and infrared light curves with the significance levels of $2\sim 2.5\sigma$. Future observations with longer durations can test and verify our results.

In addition, we obtain the $\gamma$-ray flux variations of PKS J2134-0153 from the Pass 8 database of the Fermi data server\footnote{\url{https://fermi.gsfc.nasa.gov/ssc/data/}} (see Appendix~\ref{sec:gamma}). We perform the LSP analysis of the $\gamma$-ray light curve and cannot detect any statistically significant periods. 

\begin{table*}
\caption{Observation log and parameters and results of the periodic analysis for PKS J2134-0153. \label{tab:paramenters}}
 
\centering 
\setlength{\tabcolsep}{3mm}{
\begin{tabular}{ccccccccc} 
\hline\hline 
Band & MJD & Time span (days) & $N_{\mathrm{data}}$ & $N_{\mathrm{eval}} $& $f_{\mathrm{min}}\ (\mathrm{day^{-1}})$ & $f_{\mathrm{max}}\ (\mathrm{day^{-1}})$ & P (days)& $\Delta$P (days)\\
(1) & (2) & (3) & (4) & (5) & (6)& (7)& (8) &(9)\\
\hline

Infrared (W1 \& W2) & 55332-59885 & 4553 & 162             \\
$\mathrm{W1}_{\mathrm{binned}}$ & 55332-59878 &  4546  & 19 &  90  & 1/4546  &1/505 &$1.6\times 10^3$ & $0.4\times 10^3$ \\
\hline 
V &53494-56593     & 3099       & 415       &      &      &      \\
g &58263-60244    & 1981       & 326        &         &         &          \\
r &58256-60244   & 1988       & 416       &         &         &          \\
i &58280-58771      & 491        & 42         &         &         &          \\
c &57246-60292    & 3046        & 611         &         &         &          \\
$\mathrm{V_{synthetic}}$ &53494-60292    & 6798       & 1352      & 6798 & 1/6798 &1/10  & $1.8\times 10^3$    & $1\times 10^3$     \\

\hline 
\end{tabular}}
\\
$\mathbf{Column}$ $\mathbf{notes}$: Col. (1): Band; Col. (2): Start and end time (MJD) of the observations; Col. (3): Time separation between the first and the last observation; Col. (4): Total number of observations; 
Col. (5): Total number of periodogram evaluations; Col. (6): Minimum frequency of periodogram evaluations; Col. (7): Maximum frequency of periodogram evaluations; Col. (8): Estimated period via the LSP analysis; Col. (9): Period uncertainties (half-width at half-maximum, corresponds to $\simeq 1.2 \sigma$ for a Gaussian distribution).
\end{table*}

\begin{table*}
\caption{Parameters and results of the CCF analysis for PKS J2134-0153. \label{tab:paramenters_CCF}}

\centering 
\setlength{\tabcolsep}{3mm}{
\begin{tabular}{ccccccccc}

\hline\hline 

Band & $\tau_{\mathrm{min}}$ (days) & $\tau_{\mathrm{max}}$ (days) & $T_{\mathrm{lag}}$ (days) & $\rho $ &$p$-value& $N(\mathrm{\tau}$)& ${\rho }' $ &$\Delta D$ (pc)\\
(1) & (2) & (3) & (4) & (5) & (6) & (7) & (8)&(9)\\
\hline
\specialrule{0em}{1pt}{3pt}
Infrared-Radio & $0$ & $1000$ & $\bm{(3.3\pm 2.3) \times 10^2}$ & $0.85$ & $7\times 10^{-5}$ &  $12$ &  $0.44$ & $\bm{0.37\pm 0.26}$     \\
 & $-2000$& $-500$ & $(-1.3\pm0.3)\times 10^3$ &  $0.41$ & $4\times 10^{-4}$ &  $23$ &  $0.41$ & $-1.45\pm0.33$ \\
\specialrule{0em}{1pt}{3pt}

\hline 
\specialrule{0em}{1pt}{3pt}
Optical-Radio & $0$ & $1000$ & $\bm{(3.0\pm 2.3) \times 10^2}$ &  $0.71$ & $6\times 10^{-6}$ &  $23$ &  $0.71$ & $\bm{0.33\pm 0.26}$     \\
 & $-2000$& $-500$ & $(-1.3\pm 0.4)\times 10^3$ &  $0.64$ & $0.03$ &  $23$ &  $0.64$ & $-1.45\pm 0.44$ \\

\specialrule{0em}{1pt}{3pt}
\hline
\specialrule{0em}{1pt}{3pt}
Optical-Infrared &  $-1050$  &  790  & $ \bm{0\pm140}$ &  0.81 & $1\times 10^{-5}$ &  19 & 0.81 & $\bm{0\pm0.16}$ \\

\specialrule{0em}{1pt}{3pt}
\hline
\specialrule{0em}{1pt}{3pt}
\end{tabular}}
\\
$\mathbf{Column}$ $\mathbf{notes}$: Col. (1): Band; Col. (2): Lower limit of the time lag range for determining the time lags;
Col. (3): Upper limit of the time lag range for determining the time lags; Col. (4): Time lags and its $1\sigma$ lower and upper uncertainties; Col. (5): Cross-correlation coefficients corresponding to time lags; Col. (6): $p$-value of Col. (5); Col. (7): The number of overlapping points; Col. (8): Weighted cross-correlation coefficients; Col. (9): The relative distances.
\end{table*}

\subsection{Time lag} \label{sec:timelags}
The time lags among the multi-bands of PKS J2134-0153 are calculated using the interpolated cross-correlation function \citep[ICCF; see, e.g.,][]{Peterson1998} method described in Section \ref{sec:CCF}. The adopted parameters for the PyCCF \citep[][]{Sun2018} are listed in Table \ref{tab:paramenters_CCF}. 

ON22 calculate the time lag between the infrared and the radio light curves by the cross-correlation function. They find that the infrared variation leads the radio variation by $\sim$ $250-350$ days, and the confidence level of this value is more than $2\sigma$. In this work, we re-estimate the IR-radio time lag because our IR light curves are longer than that of ON22 by about $600$ days. 

Due to the observational strategy of WISE, the gap between two adjacent epochs is about 180 days (see Section~\ref{sec:wise-obs} and Figure \ref{fig:LC}). Similarly, the 180-day binned light curves in the $V$ band (the dark red stars in the bottom panel of Figure \ref{fig:LC}) and radio band are adopted. The ICCF method requires the interpolation of unevenly sampled light curves, and the interpolation for the binned data is more statistically robust than unbinned ones. We confirm that the time-lag measurements remain unchanged if the unbinned light curves are adopted. 

\begin{figure}
\includegraphics[width=0.5 \textwidth]{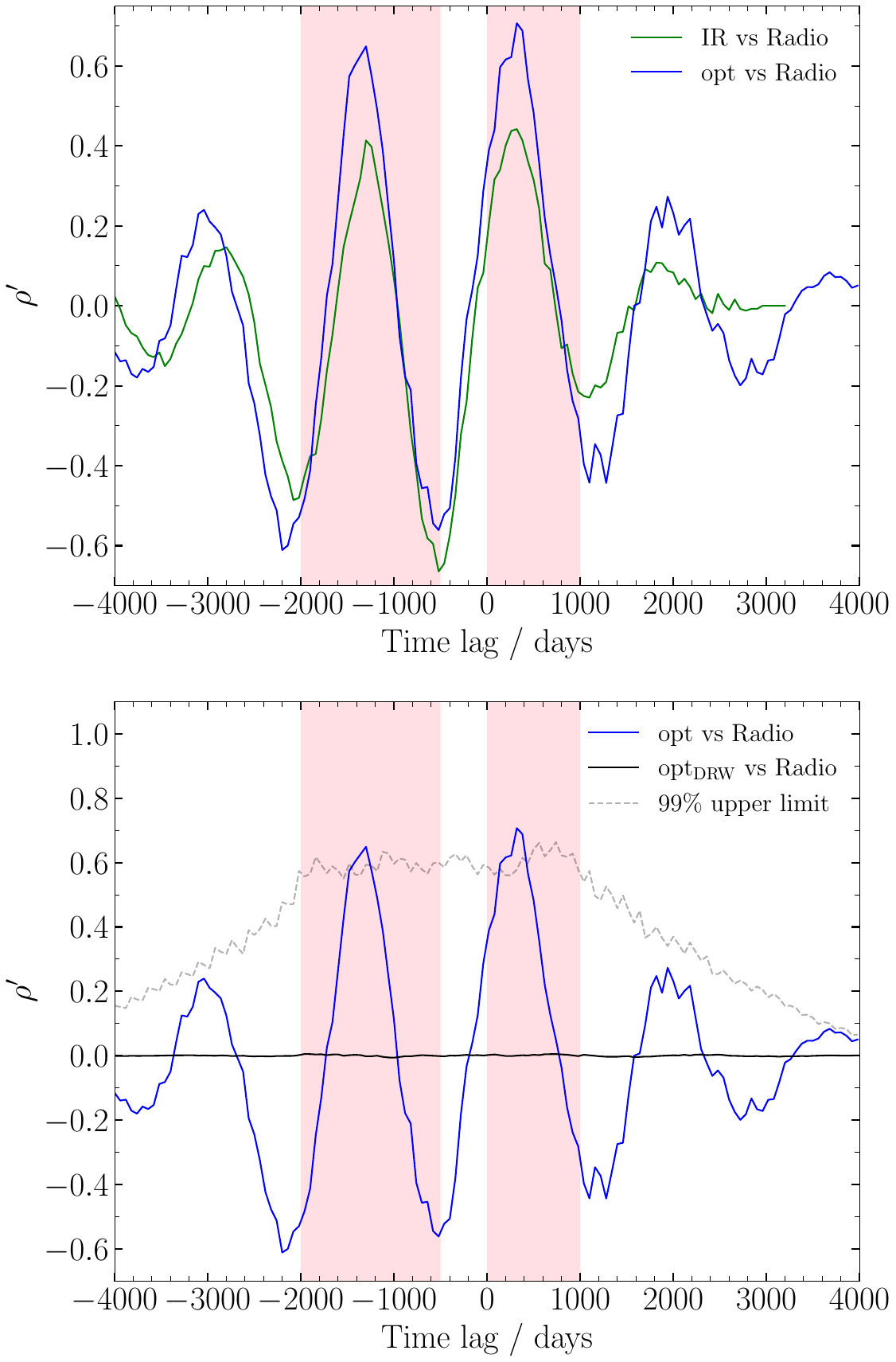}
\caption{The weighted CCFs for PKS J2134-0153. Top panel: The blue and green curves represent the CCFs for optical-radio and infrared-radio light curves, respectively. A positive value of the time lag indicates that the $V$ or infrared flux variations lead to the radio variations, and vice versa. The shaded regions indicate two CCF peaks. The re-binned $V$ and infrared light curves correlate well with the radio variations because the corresponding $p$-values are smaller than $0.01$. The time lag between the optical and radio light curves is similar to that for the infrared and radio ones. Bottom panel: the blue curve represents the CCF for optical-radio light curves. The black solid curve represents the median CCF between the optical simulated DRW light curve and the observed radio one, and the gray dashed curve represents the $99\%$ upper limit.} \label{fig:CCF}
\end{figure}

\begin{figure}
\includegraphics[width=0.48 \textwidth]{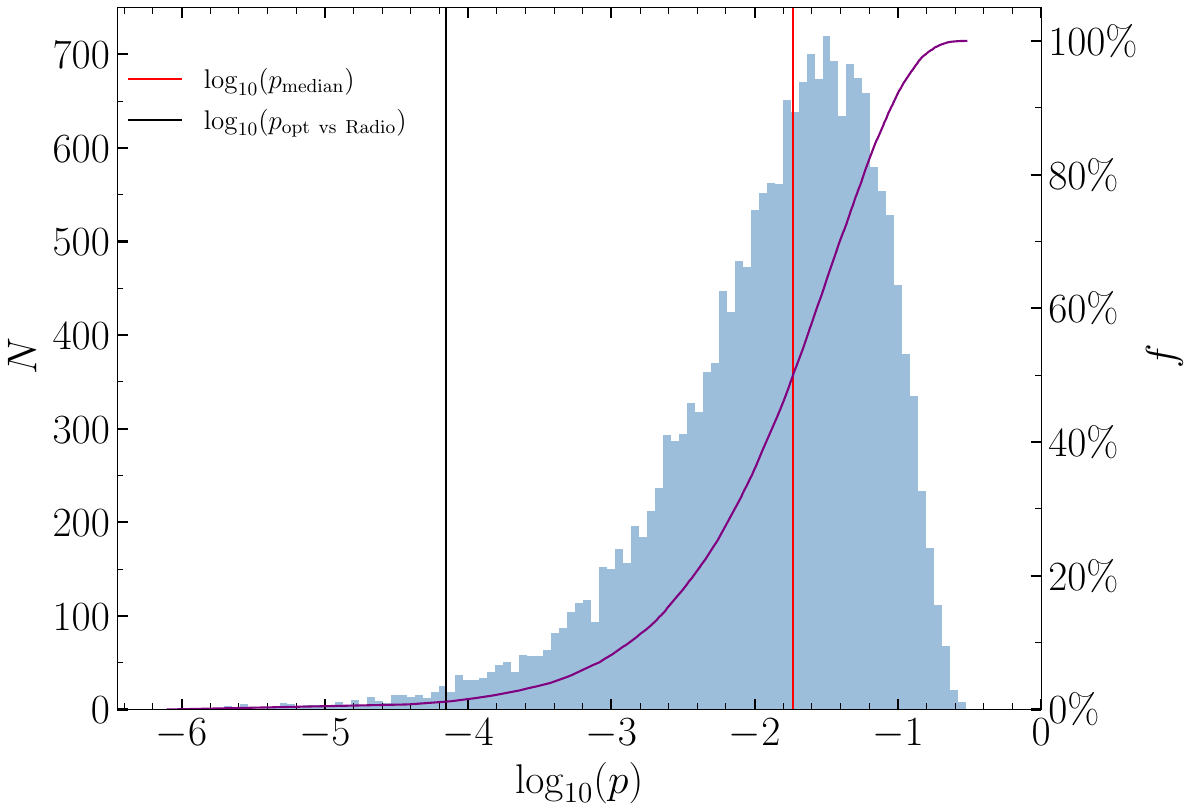}
\caption{The distribution of the logarithmic $p_{\mathrm{mock}}$ (the $p$-values corresponding to the CCF peaks between the optical simulated DRW light curves and the observed radio one). The solid purple curve represents the cumulative distribution of the logarithmic $p_{\mathrm{mock}}$. The red vertical line corresponds to the median of logarithmic $p_{\mathrm{mock}}$. The black vertical line corresponds to the logarithmic $p$-value ($p_{\mathrm{opt\ vs\ Radio}}$) of the CCF peak between the real optical and radio band light curves. The possibility of having $p_{\mathrm{mock}}<p_{\mathrm{opt\ vs\ Radio}}$ is only $0.38\%$.} \label{fig5}
\end{figure}

There are cross-correlations between the radio and optical/WISE light curves. The time lags between the 180-day binned synthetic $V$ band and the binned radio light curves are calculated from $-4000$ days to $4000$ days with the step of $\Delta \tau$ = $60$ days. High cross-correlation coefficients might be unreliable because of statistical fluctuations. Following \cite{Grier2018}, we apply a prior weight to the cross-correlation coefficients, and the weight is defined by $\omega  (\mathrm{\tau} )=\left [ N(\mathrm{\tau} )/N(\mathrm{0} ) \right ] ^{2} $, where $N(\mathrm{\tau} )$ represents the number of overlapping points at $\tau$ (see Col. (7)), and $N(\mathrm{0} )$ is the number of overlapping points at $\tau = 0$. Then, the weighted cross-correlation coefficients (${\rho }' $) can be calculated, i.e., ${\rho }' = \rho \omega $ (see Col. (8) of Table \ref{tab:paramenters_CCF}). The weighted ICCF is shown in the top panel of Figure \ref{fig:CCF}. There are two distinct peaks. For each one, we use the ICCF at the lag ranges of [$0$, $1000$] or [$-2000$, $-500$] to calculate the time lags and their uncertainties; the corresponding time lags are listed in Table~\ref{tab:paramenters_CCF}, i.e., $(3.0\pm2.3)\times 10^2$ days, and $(-1.3\pm0.4)\times10^3$ days. The cross correlation between the binned optical-radio light curves is statistically significant because the $p$-values for the two ICCF peaks are $6\times 10^{-6}$ and $0.03$, respectively. To assess the significance, we also calculate the weighted ICCF between the observed radio and DRW simulated $V$-band light curves (see Section~\ref{sec:Significance}), the results are shown in the bottom panel of Figure \ref{fig:CCF}. We simulate 20,000 optical band light curves by the DRW model, and calculate the mock ICCFs between these simulated binned optical light curves and the observed radio light curve, respectively. The gray dotted line represents the $99\%$ upper limit of the 20,000 mock ICCFs. We find that the possibility to obtain the observed ICCF peak or even higher peak is again $1\%$. We also calculate the $p$-value (hereafter $p_{\mathrm{mock}}$) for the highest peak in each mock ICCF. The distribution of the 20,000 cases of $p_{\mathrm{mock}}$ is shown as the blue histogram in Figure \ref{fig5}. The solid purple curve represents the cumulative distribution of $p_{\mathrm{mock}}$. The red vertical line corresponds to the median of these $p_{\mathrm{mock}}$. The black vertical line corresponds to the $p$-value (hereafter $p_{\mathrm{opt\ vs\ Radio}}$) of the CCF peak between the real optical and radio band light curves (the blue curve in the top panel of Figure \ref{fig:CCF}). The probability of having $p_{\mathrm{mock}}<p_{\mathrm{opt\ vs\ Radio}}$ is only $0.38\%$. Hence, we argue that the optical-radio cross-correlation is likely to be statistically significant.

\begin{figure}
\includegraphics[width=0.5 \textwidth]{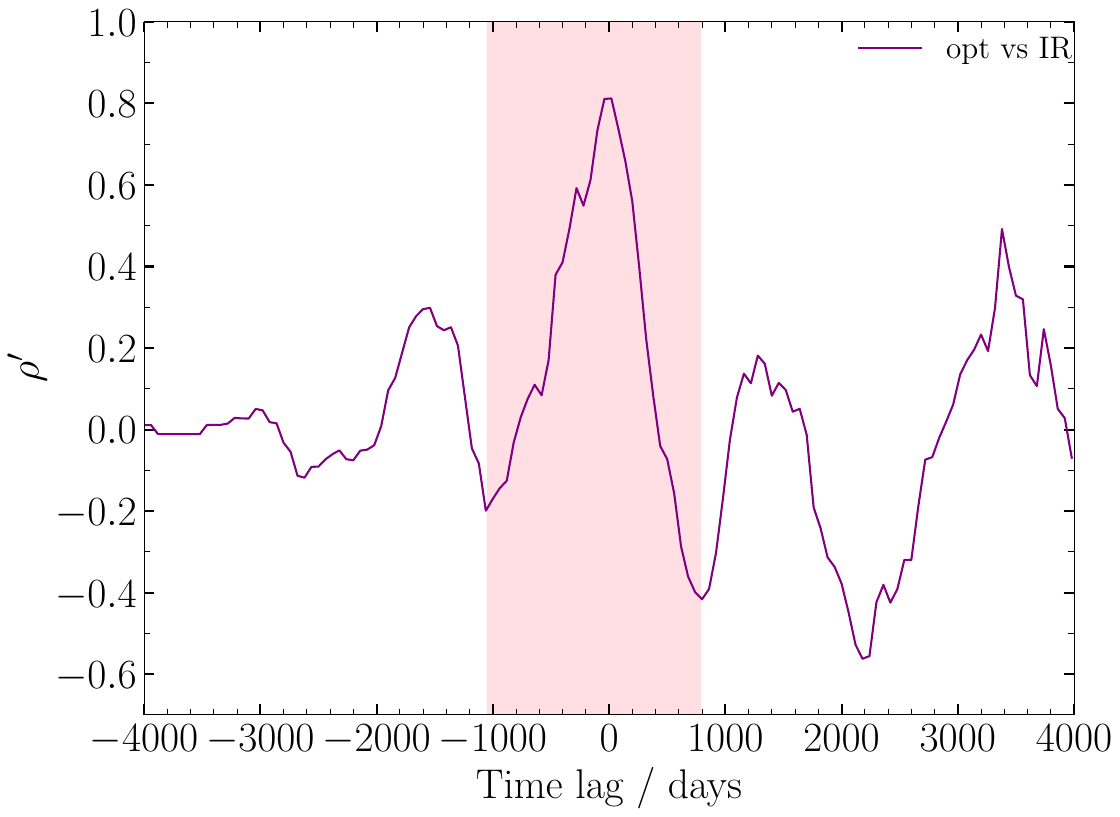}
\caption{The weighted CCFs between the binned optical and IR light curves for PKS J2134-0153.} \label{fig6}
\end{figure}

We also obtain significant cross-correlations between the radio and WISE light curves. The time lag, the maximum correlation coefficient, and the corresponding $p$-value for the WISE $W1$ band are listed in Table~\ref{tab:paramenters_CCF}. The positive peak we obtained is statistically consistent with that of ON22. Just like the radio-optical ICCF, the radio-IR ICCF also show two significant peaks and the time-lag difference for the two peaks is $\simeq 1700$ days, which is close to the variational period (see Section~\ref{sec:period}). Moreover, the radio-optical and radio-IR time lags are statistically similar. Indeed, the weighted ICCF between the binned optical-IR light curves has a high weighted cross-correlation coefficient ($\rho' = 0.81$), and the corresponding time lag is $0\pm140$ days (Figure \ref{fig6} and Table \ref{tab:paramenters_CCF}). The cross correlation between the binned optical-IR light curves is statistically significant because the $p$-value for this peak is $1\times10^{-5}$. The significant correlation with the zero-day lag suggests a common origin of the IR and optical emission. 

\section{Discussion} \label{sec:discussion}

\subsection{SMBH binary system}\label{sec:discussion_period}

We find possible periodic variations in the infrared and optical bands, and their observed-frame periods are $P_{\mathrm{IR}} = 1.6(\pm 0.4)\times 10^3$ days and $P_{\mathrm{V}} = 1.8(\pm 1)\times 10^3$ days, respectively. The two periods are statistically consistent with the period in the radio band ($P_{\mathrm{Radio}}$$ = 1760\pm33$ days; R21; ON22). Moreover, our ICCF analyses suggest that the cross-correlation between the radio and infrared/optical variations is significant. The optical period and the cross-correlation between the optical and IR emission are not reported by R21 or ON22. 

The infrared emission has two possible origins: 1) the electron synchrotron radiation in the jet \citep[e.g.,][]{Ghisellini2009, Bottcher2013}; 2) the reprocessed emission of the dusty torus. As proposed by \cite{Massaro2011}, the origin of the IR emission can be understood by the $W2-W3$ (3.22 mag) and $W1-W2$ (1.08 mag) colors where the mean magnitude of $W1$, $W2$, and $W3$ bands are $12.37$ mag, $11.29$ mag, and $8.07$ mag, respectively. Following \cite{Massaro2011} (their figure 3), the $W2-W3$ versus $W1-W2$ of our target is also consistent with being a non-thermal jet origin. 

We can estimate the rest-frame infrared and optical luminosities of this source by 
\begin{equation}
L = 4 \pi D_{\mathrm{L}}^{2}\nu f_{\nu} ,
\end{equation}
where $D_{\mathrm{L}}$ is the luminosity distance, $\nu$ is the frequency, and $f_{\nu}$ is the flux density. For the infrared bands, $f_{\nu} = f_{\nu 0} 10^{(-m/2.5)}$ and the zero-point flux $f_{\nu 0}$ for $W1$ and $W2$ \citep{Jarrett2011} are $309.54$ Jy and $171.787$ Jy, respectively. The average magnitude of the $W1$ band $m_{\mathrm{W1}} = 12.37\ \mathrm{mag}$. The rest-frame wavelength for $W1$ is $3.4\,\mu m/(1+z) \simeq 1.5\,\mu m$ and the corresponding infrared luminosity is $L_{\mathrm{1.5\,\mu \textit{m}}} = 3.1 \times 10^{46}\ \mathrm{erg\ s^{-1}}$. For the optical band, the source flux density can calculated by $m_{\nu} = -2.5\log_{10}(f_{\nu})-48.6$; the average magnitudes of $g$ and $r$ bands $m_{\mathrm{g}} = 18.07\ \mathrm{mag}$, $m_{\mathrm{r}} = 17.53\ \mathrm{mag}$, respectively; the rest-frame wavelengths for $g$ and $r$ are $4770\, \mathrm{\mathring{A}}/(1+z)=2087\,\mathrm{\mathring{A}}$ and $6231\, \mathrm{\mathring{A}}/(1+z)=2727\,\mathrm{\mathring{A}}$, respectively; the corresponding rest-frame luminosities are $L_{\mathrm{2087\,\mathring{A}}} = 1.4 \times 10^{46}\ \mathrm{erg\ s^{-1}}$ and $L_{\mathrm{2727\,\mathring{A}}} = 1.7 \times 10^{46}\ \mathrm{erg\ s^{-1}}$, respectively. The luminosity at $3000\ \mathrm{\mathring{A}}$ rest-frame ($L_{3000\,\mathrm{\mathring{A}}}= 1.8\times 10^{46}\ \mathrm{erg\ s^{-1}}$) is calculated by linearly extrapolating $L_{\mathrm{2087\,\mathring{A}}}$ and $L_{\mathrm{2727\,\mathring{A}}}$. The bolometric luminosity of an AGN disk can be estimated from $L_{\mathrm{bol}} = 5L_{3000\,\mathrm{\mathring{A}}} = 9.0\times 10^{46}\ \mathrm{erg\ s^{-1}}$ \citep[e.g.,][]{Richards2006}, which is slightly larger than $L_{\mathrm{1.5\,\mu \textit{m}}}$. It is worth noting that the optical luminosity may be underestimated if the dust extinction is important. 

Given that the infrared and optical periods are statistically consistent with the radio period, we propose that the synchrotron radiation of the same electron population produces the periodic features in the optical, infrared, and radio bands. The variation period has several possible origins. For instance, the disk-driven precession model \citep[e.g.,][]{Sarazin1980, Lu1990} due to the Lense-Thirring effect and the misalignment between the SMBH’s spin axis and the accretion disk. As a result, the jet precesses around the central supermassive black hole. The precession period can be estimated by \citep{Lu2005}
\begin{equation}
    \begin{aligned}
    \log{P}\left ( \mathrm{yr}  \right )  = &0.48M_{\mathrm{abs} }+20.06+\frac{48}{35}\log_{}{\alpha } +\frac{5}{7}\log_{}{a} \\
    &+\frac{1}{7}\log_{}{\left ( \frac{M}{10^{8}M_{\odot }  }  \right ) }+\frac{6}{5}\log_{}{\eta }       \label{equ:period},
    \end{aligned}
\end{equation}
where $M_{\mathrm{abs}}$ is the absolute magnitude of the source in the $B$ band, $\alpha$ is the dimensionless viscosity parameter, $a$ is the dimensionless specific angular momentum, $M$ is the black hole mass, and $\eta \sim 0.1$ is the radiative efficiency. $M_{\mathrm{abs}}$ can be calculated using $M_{\mathrm{abs}}=M_{\mathrm{B}}-2.5\log\left ({L_{\mathrm{B}}/L_{\odot} } \right)$, where $L_{\mathrm{B}}$ is the luminosity of the $B$ band (the rest-frame wavelength is $\sim 4400\,\mathrm{\mathring{A}}$), $M_{\mathrm{B}} = 5.44$ is the $B$ band absolute magnitude of the Sun \citep{Willmer2018}, and $L_{\odot}$ is the B band luminosity of the Sun. If we adopt $L_{\mathrm{B}} = 2.5\times 10^{46}\ \mathrm{erg\ s^{-1}}$, which is estimated by linearly extrapolating $L_{\mathrm{2087\,\mathring{A}}}$ and $L_{\mathrm{2727\,\mathring{A}}}$, and we can calculate $M_{\mathrm{abs}}= -27.31$ mag. For a typical mass of $M_{\mathrm{BH}}\simeq 10^{8}\ M_{\odot}$, $\alpha =0.4$ \citep{King2007}, and $a=0.9$ (i.e., a highly spinning SMBH), the corresponding precession period is $10^{5.2}$ years. Hence, it is hard to reconcile this precession period with the observed ones in optical and IR, unless the spin is close to zero, $L_{\mathrm{B}}\simeq 1.1\times10^4L_{\mathrm{1.5\,\mu \textit{m}}}$, or the dimensionless viscosity parameter $\alpha$ is\footnote{Such an extreme requirement of $\alpha$ has been pointed out by \cite{Sarazin1980} (see their Eq.~7).} $\simeq 10^{-4}$. The second class of models involves an SMBH binary. In such a system, the precession of the jet can be induced because of the misalignment of the orbital plane of the secondary SMBH with the primary SMBH's accretion disk \citep[e.g.,][]{Abraham2000, Romero2000, Caproni2004, Caproni2017, De2004, Liu2007, Serafinelli2020, Witt2022, Casey2022} or the accretion disk warping  \citep[e.g.,][]{Britzen2018}. Hence, the angle between the jet and our sightline changes periodically on the orbital timescale. The same mechanism is one of the physical explanations for the geometric helical jet that explains the periodic variations in a number of blazars \citep[e.g.,][]{Conway1993, Bahcall1995, Tateyama1998, Zhou2018, Jorstad2022, Zhang2022a}. Based on the long-term ($\sim 45$ yrs in observed frame) radio variations, ON22 propose that PKS J2134-0153 hosts a binary SMBH and the orbital motion of the primary SMBH (along with the jet) should be responsible for the radio light curve, and the separation of the two SMBHs is only about $10^{16}$ cm (also see R21). In this work, we find that the infrared and optical periods are statistically consistent with the radio period. In addition, there is a significant correlation between the $V$ and the infrared emission with significant time lags. These new results suggest that the driving mechanism for the periodicity for the optical and IR emission regions is similar to the radio emission region, albeit there is about $\sim 1$ pc separation between the optical/IR and radio emission regions. That is, the jet precession acts as a rigid body. In summary, our periodicity and cross-correlation analyses reveal more details about the jet precession in the SMBH binary candidate PKS J2134-0153 (e.g., R21; ON22).

\subsection{The position of emission regions \label{sec:position}}

As shown in Section \ref{sec:period}, the periodic variations in the infrared band and optical band ($P_{\mathrm{IR}} = 1.6(\pm 0.4)\times 10^3$ days, $P_{\mathrm{V}} = 1.8(\pm 1)\times 10^3$ days) are statistically consistent with the radio period. The significance level of $P_{\mathrm{V}}$ is more than 2.5$\sigma$. ON22 reported a strong correlation (with a correlation coefficient $\sim$1) between the infrared band and the radio band light curves. We find the significant cross-correlations between infrared, optical, and radio variations since the $p$-values are much smaller than $0.01$ (see Section \ref{sec:timelags}). In summary, the infrared, radio, and a significant fraction of optical variable fluxes are likely to be produced by the same precession or helical jet. 

The time lags in optical, infrared, and radio variations suggest that the emission regions of optical/infrared are different from radio emission regions. The optical and infrared emission regions may be overlapped significantly. \cite{Max-Moerbeck2014} proposed a model to interpret the time lags. In their model, a moving disturbance moves across the jet core and induces multi-band variations in their corresponding emission regions; the infrared emission regions are closer to the center SMBH than the radio emission regions, and the $V$ band emission regions should be closer to the center SMBH than the infrared band. Hence, the negative time lags should be excluded. In the subsequent analysis, we adopt $(3.3\pm 2.3) \times 10^2$ days as the time lag between infrared and radio emission, and $(3.0\pm 2.3) \times 10^2$ days as the time lag between optical and radio emission. The relative distance between different emission regions can be calculated by the time lags between the corresponding light curves, i.e., 
\begin{equation}
    \Delta\mathrm{D} =  \frac{\Gamma \delta \beta c T_{\mathrm{lag}}}{\left ( 1+z \right ) } , \label{equ:D}
\end{equation}
where $\Gamma$ is the bulk Lorentz factor, $\delta$ is the Doppler factor, $\beta c$ is the bulk jet speed, $c$ represents the speed of light, $T_{\mathrm{lag}}$ is the time lag and $z$ is the redshift \citep[e.g.,][]{Pushkarev2010, Max-Moerbeck2014}. Following \cite{Pushkarev2010}, we can use the $\delta$-$\beta_{\mathrm{app}}$ relation of \cite{Cohen2007} and Eq.\ref{equ:D} to obtain
\begin{equation}
\Delta D = \frac{\beta_{\mathrm{app}}cT_{\mathrm{lag}}}{\sin \theta (1+z)}, \\ \label{equ:delta D}
\end{equation}
where $\beta_{\mathrm{app}}$ and $\theta$ are the apparent bulk jet speed and the viewing angle, respectively. 
We obtained the apparent bulk jet speed $\beta_{\mathrm{app}}=0.19\pm 0.12$ from the MOJAVE (Monitoring Of Jets in Active galactic nuclei with Very Long Baseline Interferometry (VLBI) Experiments) \citep[e.g.,][]{Lister2005, Lister2019}. The viewing angle is assumed to be $3^{\circ}.6$, a typical value of blazars \citep[][]{Pushkarev2009, Hovatta2009}. The time lags among different bands are listed in Table \ref{tab:paramenters_CCF}. Hence, the relative distances between the infrared and radio regions and between the optical and radio regions are $0.37\pm0.26$ pc and $0.33\pm0.26$ pc, respectively.

The radio-core size can determined by the angular diameter, $\theta_{\mathrm{core}}$,
\begin{equation}
    d_{\mathrm{core}} \sim  \frac{\left ( \theta_{\mathrm{core}}/2 \right ) d_{\mathrm{A} }  }{\tan \left ( \alpha _{\mathrm{int} }/2 \right ) }, \label{equ:D_core}
\end{equation}
where $d_{\mathrm{A}}=1742.7\;\mathrm{Mpc}$ is the angular diameter distance for a $\Lambda$ cold dark matter cosmology \citep[e.g.,][]{Komatsu2011}, and redshift $z=1.285$. $\theta_{\mathrm{core}}$ represents the angular diameter of the radio core, $\alpha _{\mathrm{int}}$ is the intrinsic opening angle. For PKS J2134-0153, $\theta_{\mathrm{core}}$ is measured by the MOJAVE \citep[][]{Lister2005, Lister2019}, which is $\theta_{\mathrm{core}} = 0.3\;\mathrm{mas}$ \citep[][]{Kovalev2005}. We adopt $\alpha _{\mathrm{int}} = 2^{^\circ}.4\pm2^{^\circ}.0$, the mean value for BL Lacs \citep[][]{Pushkarev2009}. According to Equation \ref{equ:D_core}, we find $d_{\mathrm{core}}=60\pm11\;\mathrm{pc}$. The relative distances between the infrared and radio and between the optical and radio are much smaller than the core size.

These size measurements can be used to test jet models. For the synchrotron self-absorption model \citep[for details, see][]{Lobanov1998, Kovalev2008a, Kovalev2008b, OSullivan2009, Tramacere2022}, the positions of the different emission regions scale as 
\begin{equation}
    r(\mathrm{\nu} )\propto \nu ^{-1} , \label{equ:r_core}
\end{equation}
where $r$ and $\nu$ are the position of the emission region and the frequency, respectively. According to Equation~\ref{equ:r_core}, the distances between the different emission regions can be calculated 
\begin{equation}
    \Delta r_{ \mathrm{IR-Radio}}^{\mathrm{th}} =r_{\mathrm{Radio} }\left ( 1 - \frac{\nu_{\mathrm{Radio} } }{\nu_{\mathrm{IR} }} \right ) ,  \label{equ:r_Radio-IR}
\end{equation}

\begin{equation}
    \Delta r_{\mathrm{V-Radio}}^{\mathrm{th}} =r_{\mathrm{Radio} }\left ( 1-\frac{\nu_{\mathrm{Radio} } }{\nu_{\mathrm{V} }}  \right ) , \label{equ:r_V-IR}
\end{equation}
where $r_{\mathrm{Radio} }$ is the absolute position of the radio band emission region. Hence, their ratio is $f_\mathrm{th} = \Delta r_{ \mathrm{V-Radio}}^{\mathrm{th}} / \Delta r_{ \mathrm{IR-Radio}}^{\mathrm{th}}=(1-\nu_{\mathrm{Radio}}/\nu_{\mathrm{V}}) / (1 - \nu_{\mathrm{Radio}}/\nu_{\mathrm{IR}}) \simeq 1$. This ratio is consistent with the observed one ($0.9\pm 0.7$), and also consistent with the $0$-day time lag between the optical and infrared band light curves.

\section{Conclusions} \label{sec:conclusion}

In this paper, we have systematically searched possible periodic variations in the optical and infrared light curves of PKS J34-0153. Our results are as follows:

(1) Using the Lomb-Scargle periodogram, we have found a periodic signal with a possible period of $1.6(\pm 0.4)\times 10^3$ days in the infrared band light curve and a $1.8(\pm 1)\times 10^3$ days possible period in the $V$ band of PKS J2134-0153 (see Section \ref{sec:period}, or Figure \ref{fig:LSP}). The two periods (both in the observed frame) are consistent with the period in the radio band ($P_{\mathrm{Radio}} = 1760 \pm 33$ days, R21; ON22). A similar period was recently reported by \cite{Kiehlmann2024}. The SMBH binary system likely generates the periodic variations in these three bands. 

(2) The distances between various emission regions are calculated according to the inter-band time lags. The relative distance between the infrared and radio emission regions is $0.37\pm0.26$ pc. The relative distance between the V and radio emission regions is $0.33\pm 0.26$ pc (see Section \ref{sec:position} and Table \ref{tab:paramenters_CCF}). The time lag between the infrared and V band light curves is $0\pm140$ days (see Section \ref{sec:timelags}, Table \ref{tab:paramenters_CCF}, and Figure \ref{fig6}). This suggests that the emission regions of the infrared and optical bands may be very close to each other. The relative distances are quantitatively consistent with the theoretical prediction \citep[][]{Lobanov1998, Kovalev2008a, Kovalev2008b, OSullivan2009, Tramacere2022}.

The long multi-band monitoring of PKS J2134-0153 is needed to confirm our conclusions in future work.

\section{Acknowledgment}
G.W.R. and M.Y.S. acknowledge support from the National Natural Science Foundation of China (NSFC-11973002), the Natural Science Foundation of Fujian Province of China (No. 2022J06002), and the China Manned Space Project grants (No. CMS-CSST-2021-A06; No. CMS-CSST-2021-B11). N. D. is sincerely grateful for the financial support of the Xingdian Talents Support Program, Yunnan Province (NO. XDYC-QNRC-2022-0613). Z.X.Z acknowledges support from the National Science Foundation of China (NSFC-12033006; NSFC-12103041). We thank Haikun Li for maintaining the computational resources. We thank the referee for his/her constructive comments that greatly improved the manuscript. 

The CSS survey is funded by the National Aeronautics and Space Administration under Grant No. NNG05GF22G issued through the Science Mission Directorate Near-Earth Objects Observations Program.  The CRTS survey is supported by the U.S.~National Science Foundation under grants AST-0909182.

Based on observations obtained with the Samuel Oschin Telescope 48-inch and the 60-inch Telescope at the Palomar Observatory as part of the Zwicky Transient Facility project. ZTF is supported by the National Science Foundation under Grants No. AST-1440341 and AST-2034437 and a collaboration including current partners Caltech, IPAC, the Weizmann Institute for Science, the Oskar Klein Center at Stockholm University, the University of Maryland, Deutsches Elektronen-Synchrotron and Humboldt University, the TANGO Consortium of Taiwan, the University of Wisconsin at Milwaukee, Trinity College Dublin, Lawrence Livermore National Laboratories, IN2P3, University of Warwick, Ruhr University Bochum, Northwestern University and former partners the University of Washington, Los Alamos National Laboratories, and Lawrence Berkeley National Laboratories. Operations are conducted by COO, IPAC, and UW.

This work has made use of data from the Asteroid Terrestrial-impact Last Alert System (ATLAS) project. The Asteroid Terrestrial-impact Last Alert System (ATLAS) project is primarily funded to search for near earth asteroids through NASA grants NN12AR55G, 80NSSC18K0284, and 80NSSC18K1575; byproducts of the NEO search include images and catalogs from the survey area. This work was partially funded by Kepler/K2 grant J1944/80NSSC19K0112 and HST GO-15889, and STFC grants ST/T000198/1 and ST/S006109/1. The ATLAS science products have been made possible through the contributions of the University of Hawaii Institute for Astronomy, the Queen’s University Belfast, the Space Telescope Science Institute, the South African Astronomical Observatory, and The Millennium Institute of Astrophysics (MAS), Chile.

This publication makes use of data products from the Near-Earth Object Wide-field Infrared Survey Explorer (NEOWISE), which is a joint project of the Jet Propulsion Laboratory/California Institute of Technology and the University of Arizona. NEOWISE is funded by the National Aeronautics and Space Administration.

SDSS-IV is managed by the Astrophysical Research Consortium for the Participating Institutions of the SDSS Collaboration including the Brazilian Participation Group, the Carnegie Institution for Science, Carnegie Mellon University, the Chilean Participation Group, the French Participation Group, Harvard-Smithsonian Center for Astrophysics, Instituto de Astrof{\'i}sica de Canarias, The Johns Hopkins University, Kavli Institute for the Physics and Mathematics of the Universe (IPMU)/University of Tokyo, the Korean Participation Group, Lawrence Berkeley National Laboratory, Leibniz Institut f{\"u}r Astrophysik Potsdam (AIP), Max-Planck-Institut f{\"u}r Astronomie (MPIA Heidelberg), Max-Planck-Institut f{\"u}r Astrophysik (MPA Garching), Max-Planck-Institut f{\"u}r Extraterrestrische Physik (MPE), National Astronomical Observatories of China, New Mexico State University, New York University, University of Notre Dame, Observat{\'a}rio Nacional/MCTI, The Ohio State University, Pennsylvania State University, Shanghai Astronomical Observatory, United Kingdom Participation Group, Universidad Nacional Aut{\'o}noma de M{\'e}xico, University of Arizona, University of Colorado Boulder, University of Oxford, University of Portsmouth, University of Utah, University of Virginia, University of Washington, University of Wisconsin, Vanderbilt University, and Yale University.

\section{Data Availability}
\label{sec:download}
The CSS light curve can be downloaded from \url{https://crts.caltech.edu/}; the ZTF and WISE light curves can be requested via \url{https://irsa.ipac.caltech.edu}; the ATLAS observations can be obtained from \url{https://atlas.fallingstar.com/}. The code and data are available in its online supplementary material.






\appendix
\section{The $\gamma$-ray light curve of PKS J2134-0153} \label{sec:gamma}

\begin{figure*}
\includegraphics[width=1 \textwidth]{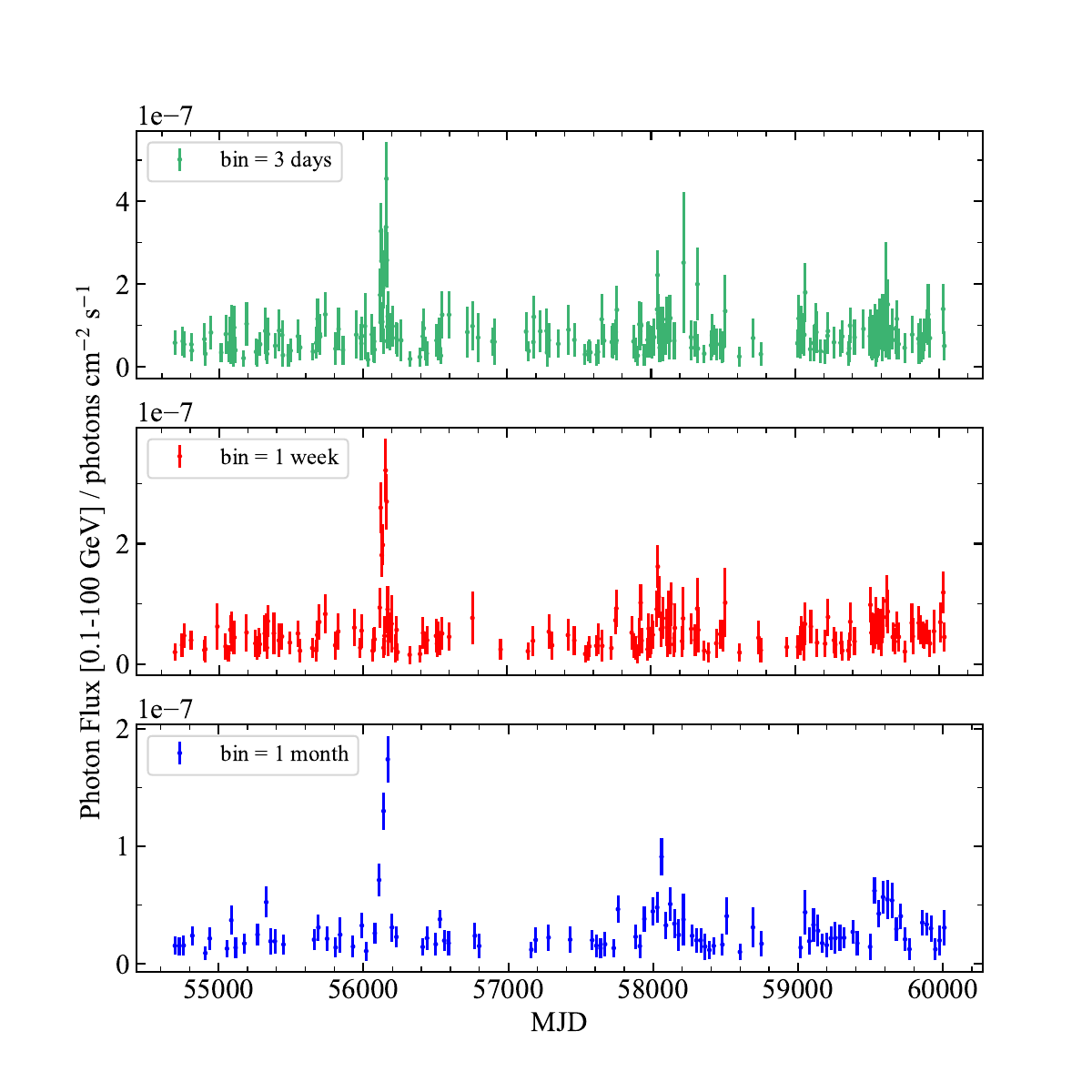}
\caption{The 0.1 \textendash 100 GeV $\gamma$-ray light curves of PKS J2134-0153 which are rebinned with 3 days (upper), 7 days (middle), and 30 days (lower).   \label{fig:LC_gamma}}
\end{figure*}

The Fermi $\gamma$-ray Space Telescope was launched in 2008. One of its instruments is the Large Area Telescope \citep[LAT,][]{Atwood2009}, which made long-term monitoring of the $\gamma$-ray sky over a wide range of $\gamma$-ray energy ranges. The Fermi LAT Light Curve Repository \citep[LCR;][]{Valverde2022}\footnote{\url{https://fermi.gsfc.nasa.gov/ssc/data/access/lat/LightCurveRepository/}} contains a public database of variable Fermi-LAT sources' light curves on various timescales. The 0.1 \textendash 100 GeV $\gamma$-ray flux variations of PKS J2134-0153 are obtained from LCR. The light curves are rebinned with 3 days, 7 days (weekly), and 30 days (monthly), as shown in Figure \ref{fig:LC_gamma}. The start and end times of the three light curves are from MJD 54691 to MJD 60038 (5347 days). 

\section{The construction of the synthetic $V$-band light curve} \label{sec:app}
\begin{figure*}
\includegraphics[width=1 \textwidth]{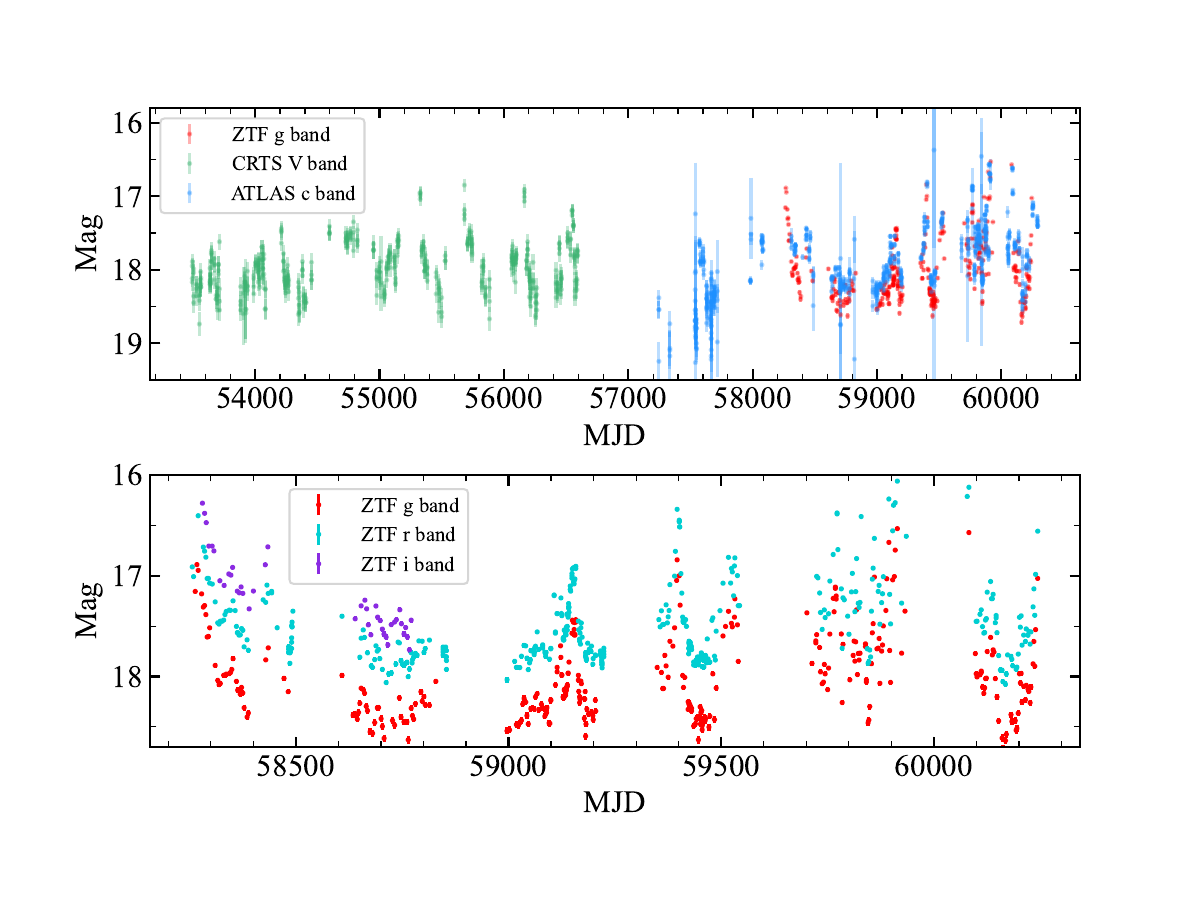}
\caption{The original optical light curves of PKS J2134-0153. Top panel: the ZTF-$g$ band, CRTS-$V$ band, and ATLAS-$c$ band original light curves. Bottom panel: the $g$, $r$, and $i$ band light curves of PKS J2134-0153 obtained from ZTF. \label{fig:LC_optical}}
\end{figure*}

\begin{figure*}
\includegraphics[width=0.8 \textwidth]{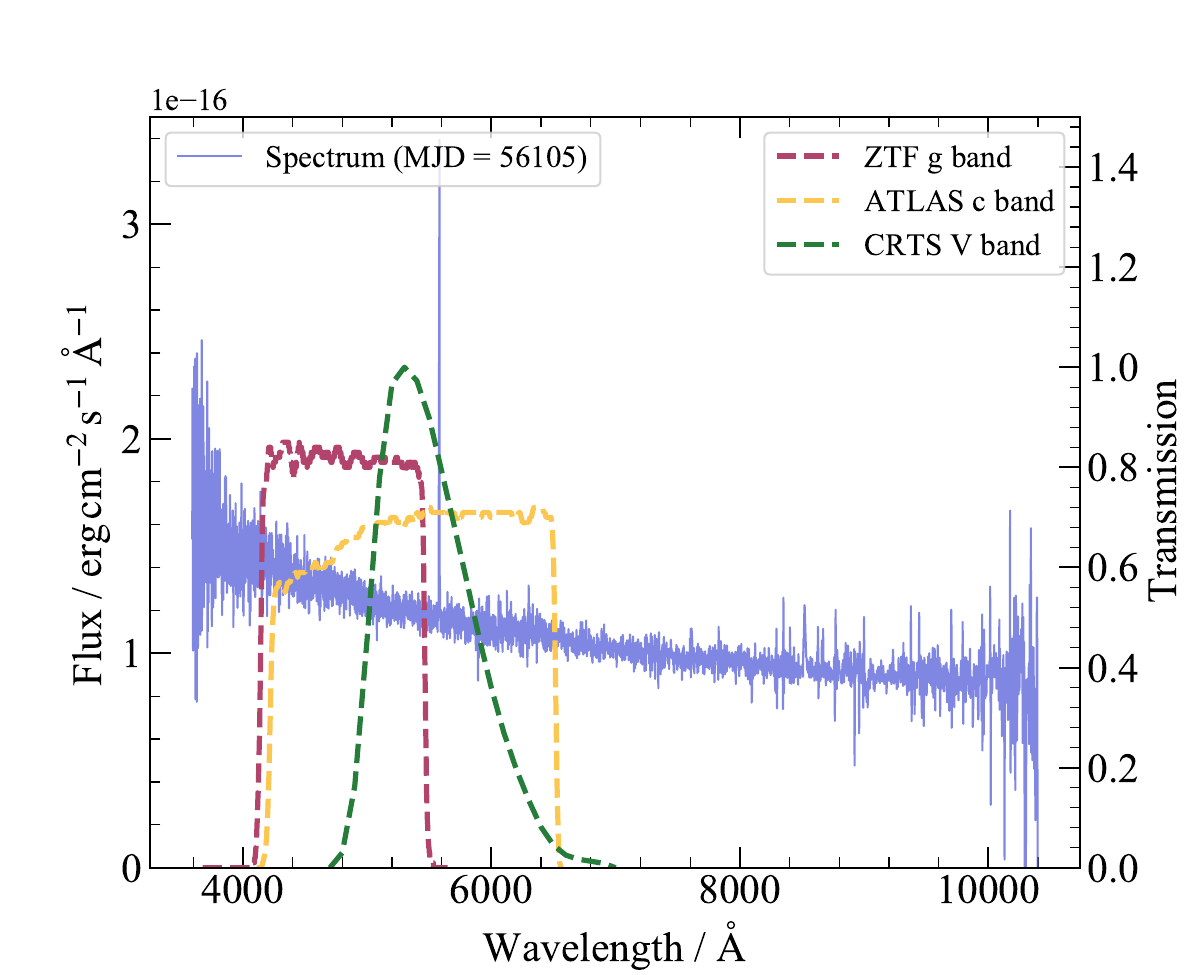}
\caption{The optical spectrum (plate-mjd-fiberid=4384-56105-0988) of PKS J2134-0153 (blue curve). The filter response curves of CRTS $V$ band (green curve), ZTF $g$ band (red curve), and ATLAS $c$ band (yellow curve).  \label{fig:spectra}}
\end{figure*}

The CRTS, ATLAS, and ZTF light curves (Figure~\ref{fig:LC_optical}) are obtained by different telescopes with different filters. We aim to merge them to construct a $\sim 6000$-day (observed frame) long optical light curve. Hence, we must correct for the filter differences. Our approach is to convolve the SDSS spectrum of PKS J2134-0153 (Figure~\ref{fig:spectra}) with the filter response curves and obtain the offsets between CRTS-$V$, ZTF-$g$, and ATLAS-$c$ (as explained in Section~\ref{sec:optical}). We use these three filters because their effective wavelengths are close to each other. Our approach is based on the implicit assumption that the optical color of PKS J2134-0153 varies weakly. Indeed, the ZTF $g-r$ color of PKS J2134-0153 only changes by $\sim 0.05$ mag, much smaller than the flux-variation amplitudes. 

\bsp	
\label{lastpage}
\end{document}